\documentclass[conference,compsoc]{IEEEtran}
\pdfoutput=1
\usepackage{multirow}
\usepackage{hhline}

\ifCLASSOPTIONcompsoc

\usepackage[nocompress]{cite}
\else
\usepackage{cite}
\fi

\usepackage{makecell}
\usepackage{xcolor}

\usepackage{tabularx}
\usepackage{graphicx}
\usepackage[capitalise]{cleveref}
\usepackage[font={small}]{caption}
\usepackage{url}

\begin{document}

	\title{\textbf{Memory Controller Design Under Cloud Workloads}}
	\author{\IEEEauthorblockN{Mostafa Mahmoud}
		\IEEEauthorblockA{Electrical and Computer Engineering\\
			University of Toronto\\
			Toronto, ON, Canada\\
			mostafa.mahmoud@mail.utoronto.ca}
		\and
		\IEEEauthorblockN{Andreas Moshovos}
		\IEEEauthorblockA{Electrical and Computer Engineering\\
			University of Toronto\\
			Toronto, ON, Canada\\
			moshovos@eecg.toronto.edu}
		}
	
	\newcommand{\fixme}[1]{\textcolor{red}{#1}}

	\maketitle

	\begin{abstract}
    This work studies the behavior of state-of-the-art memory controller designs when executing scale-out workloads. It considers memory scheduling techniques, memory page management policies, the number of memory channels, and the address mapping scheme used.  Experimental measurements demonstrate: 1)~Several recently proposed memory scheduling policies are not a good match for these scale-out workloads. 2)~The relatively simple First-Ready-First-Come-First-Served (FR-FCFS) policy performs consistently better, and 3)~for most of the studied workloads, the even simpler First-Come-First-Served scheduling policy is within 1\% of FR-FCFS. 4)~Increasing the number of memory channels offers negligible performance benefits, e.g., performance improves by 1.7\% on average for 4-channels vs. 1-channel. 5)~77\%-90\% of DRAM rows activations are accessed only once before closure. These observation can guide future development and optimization of memory controllers for scale-out workloads.
	\end{abstract}

	\section{Introduction}
	Today there is an increasing demand for cloud services, such as media streaming, social networks and search engines. Cloud services providers have been expanding their computing infrastructure adding more data centers comprising many computing systems. The performance and power consumption of these systems dictates how capable these data centers are. Accordingly, improving the performance and power of these systems can greatly improve overall data center efficiency. The first step in improving these systems is understanding their behavior in order to identify inefficiencies and opportunities for improvement. 
	
    The majority of modern server processors are built using the same microarchitecture as the one used for consumer designs with minor tweaks. There is already evidence suggesting that these architectures are suboptimal and that simpler processors tend to perform better for scale-out workloads~\cite{PicoServer:Kgil,search_Mob_Cores,scale_outProc}. 
Ferdman \textit{et al.} did an extensive study and instrumentation of \textit{CloudSuite}, a representative set of cloud applications~\cite{clearing_clouds:babak} and discovered significant over-provisions, at the level of microarchitectural structures and cache and memory hierarchy, in currently widely used server processors. Their study focused on microarchitectural characteristics and identified several inefficiencies including long instruction-fetch stalls, under-utilized instruction-level parallelism (ILP) and memory-level parallelism (MLP) structures, last-level cache (LLC) ineffectiveness, over-sized L2 cache, reorder buffer and load-store queue under-utilization and off-chip bandwidth over-provisioning. Our work builds upon this past study and focuses on the memory controller, a component that has not been covered in detail yet.

Specifically, this work studies the implications of scale-out workloads characteristics on memory controller design aspects including memory scheduling techniques, DRAM page management policies, the number of memory channels, and the indexing scheme used to access memory channels. Over the last decade, researchers and processor vendors have been developing even more sophisticated on-chip memory controllers that consider a wide range of system status attributes in scheduling decisions \cite{FR_FCFS:Rixner,ATLAS:Mutlu,PAR_BS:Mutlu,FQM,STFM:Mutlu,RL:Mutlu}. Furthermore, there has been a trend to increase the number of memory channels with four on-chip channels being available on high-end designs \cite{intel_xeon,amd_opteron}. Increasing the number of memory channels offers more parallelism and allows more memory to be connected to the same processor chip.
	
	This work characterizes the performance of several state-of-the-art memory controller designs when running several modern server workloads. The emphasis is on the CloudSuite benchmarks which are the best available to us workloads representing modern cloud applications. This study also considers traditional server workloads from SPECweb99, online transaction processing (OLTP) and decision support applications for completeness and comparison purposes. The following memory scheduling algorithms are studied: ATLAS \cite{ATLAS:Mutlu}, PAR-BS \cite{PAR_BS:Mutlu} and Reinforcement Learning \cite{RL:Mutlu}. Section \ref{Scheduling Algorithms} reviews these policies.
    
    The following key conclusions are drawn:
	\begin{itemize}
		\item The First-Ready First-Come-First-Served (FR-FCFS) algorithm \cite{FR_FCFS:Rixner} outperforms the other state-of-the-art memory scheduling algorithms for all the studied server workload categories. These other techniques were designed for desktop and scientific applications with abundant memory level parallelism (MLP) or for multi-programmed heterogeneous memory-intensity workloads. The server workloads studied do not exhibit these characteristics, and thus memory controller design ought to be revisited to better match these workloads.
		\item For five out of the six studied scale-out workloads, the performance of the simpler First-Come-First-Served (FCFS) memory scheduling algorithm is within 1\% of the baseline FR-FCFS. Accordingly, using this simpler policy may be best in some cases.
		\item Out of all DRAM row activations, 77\%-90\% receive only a single access before they are being forced to close. Leaving a row open till a conflict happens increases memory access latency for subsequent requests whereas proactively closing a row that will not receive any further hits reduces it. Accordingly, smarter page management policies may improve overall memory access latency. 
        \item State-of-the-art page management policies degrade performance by 4\% for scale-out workloads but did improve performance by 3\% for decision support workloads. 
		\item A multi-channel memory controller does not improve the performance for scale-out workloads but does benefit decision support workloads. Specifically, decision support workloads performance improves on average by 19\% on a 4-channel system compared to a single-channel one. 
	\end{itemize}
	
    Overall, the results of this study emphasize the unique off-chip memory access demands of scale-out workloads and suggest that simpler memory controllers specialized for cloud workloads may be beneficial. A limitation of this study is that it does not directly consider energy and power consumption focusing primarily on performance. Future work may address this limitation. However, the results of this work are valuable as the techniques that perform best at the end are also the simplest to implement and hence would also reduce overall energy and power consumption.
	The rest of this paper is organized  as follows.  Section \ref{Background} reviews state-of-the-art memory controller designs including memory scheduling techniques and page management policies. Section \ref{Methodology} presents the experimental methodology. Section \ref{results} reports our findings. Section \ref{limitations} discusses some of the limitations of this study. Section \ref{related work} summarizes related work and Section \ref{conclusion} concludes.
	
	\section{Background} \label{Background}
	This section reviews the two core memory controller  components that this study characterizes: the scheduling algorithm (Section \ref{Scheduling Algorithms}) and the memory page management policy, or \textit{page management policy} (Section \ref{page policies}). The former decides which memory request to service next while the latter decides  after servicing a request, whether to keep the row open or precharge it, indirectly affecting the service time of any subsequent requests.  
	
	\subsection{Memory Scheduling Algorithms} \label{Scheduling Algorithms}
    A memory scheduling algorithm (MSA) specifies how the memory controller decides the next request to service given a pool of outstanding requests. The MSA's inputs are at the very least the state of the DRAM banks and buses, and a pool of waiting requests. Additional statistics or system attributes may also be used to guide the decision making. MSAs can target different objectives such as memory throughput or fairness among running threads. The MSA affects the waiting time each request experiences before being serviced.
    Since main memory is a shared resource, requests from all cores contend and the MSA affects overall system and individual thread performance.
	
	The FCFS scheduling algorithm services requests in the order they arrive at the memory controller. FCFS does not exploit row-buffer locality as it does not try to reorder requests to increase the row-buffer hit rate. If cores generate memory requests at different rates, cores with a low memory-intensity access stream can suffer from starvation and long memory access latencies. FCFS is the simplest memory scheduling technique and its hardware overhead and power consumption are the lowest among those we study. We evaluate FCFS\_banks, a variant of FCFS that maintains separate per-bank request queues and thus can exploit bank-level parallelism.
	
	FR-FCFS \cite{FR_FCFS:Rixner} separates requests into two groups depending on whether they will hit in the currently open row, and further, within each group it orders requests according to age. It prioritizes hits over misses, and older over younger requests. FR-FCFS's goal is to increase memory throughput. FR-FCFS can lead to starvation and low overall throughput when there is imbalance in the row-buffer locality of the access streams across cores~\cite{denial_mem:mutlu,STFM:Mutlu,PAR_BS:Mutlu}. 
	
	Parallelism-Aware Batch Scheduling (PAR-BS) \cite{PAR_BS:Mutlu} targets maintaining fairness among cores and reducing the average stall time through \textit{Batching} and \textit{Ranking}. Batching groups the oldest $N$ requests from each core into a batch that is prioritized over the remaining requests. After batch formation, the cores within the batch are ranked using a shortest-job first criterion where the core with the minimum number of requests to any bank is considered the core with the shortest job. Ranking minimizes the average waiting time across all cores. 
	
	Adaptive per-Thread Least-Attained-Service memory scheduling (ATLAS) \cite{ATLAS:Mutlu} is based on the observation that cores that require less memory service time are more vulnerable to interference from cores that require more. ATLAS divides time into 10M cycle quantums tracking the total memory service time used by each core during each quantum as follows: every cycle, the attained service time (ATS) of a core is increased by the number of banks servicing its requests. At the start of the subsequent quantum, the memory controller ranks the cores according to their ATS, ranking those with less ATS higher. This policy exploits bank-level parallelism and reduces average latency across cores. ATLAS prevents starvation by prioritizing requests that have been waiting for more than a threshold $T$ cycles.
	
	Reinforcement Learning-based memory scheduling algorithm (RL) \cite{RL:Mutlu} uses reinforcement learning \cite{RLBook:Sutton} resulting in a self-optimizing memory scheduler. RL uses a number of attributes to represent the current system \textit{state (s)} such as the number of reads in the request queue, the number of writes and the number of load requests. The \textit{action (a)} that it can perform at a given cycle is precharge, activate, write, read for a load miss, read for a store miss or no-action. For every possible system state-action pair $(s, a)$, RL associates a \textit{Q-value} that represents the expected future reward if action $a$ is executed in state $s$. Given a current state, RL's goal is to maximize the future rewards by choosing the action with the highest \textit{Q-value}. RL uses an area and energy efficient implementation of the aforementioned reinforcement learning algorithm. With a preset probability, the algorithm decides to execute a random action instead of the one with the highest \textit{Q-value} to explore new search space areas.
	
	The Fair Queuing Memory Scheduler (FQM)~\cite{FQM} is a memory scheduling algorithm based on a computer network fair queuing algorithm. FQM's goal is to provide equal memory bandwidth to all cores. In FQM, each bank in the DRAM keeps a \textit{virtual time} counter per core that is increased every time a request from the corresponding core is serviced by that bank. Each bank prioritizes the core with the earliest virtual time since this is the core that got the least service from that bank. Later proposals outperformed FQM~\cite{PAR_BS:Mutlu,STFM:Mutlu} as the latter does not consider the long term intensity of the memory access stream per core and does not attempt to maximize row hits. For this reason, we do not consider it further.
	
    
	\subsection{Page Management Policies} \label{page policies}
	The page management policy decides for how long a memory row or \textit{page} should remain open. One such policy is the open-page management policy (O$_{PM}$) where a row will be closed only if another request forces it to do so. The O$_{PM}$ tends to be a good match for single-core systems whose memory access stream often exhibits good spatial locality. This way, subsequent requests  hit in the row-buffer and thus experience lower latency. 
	
    The memory controller of many-core systems observes the interleaving of several access streams and as such may not exhibit sufficient spatial locality. In these systems, row hits are less likely and waiting to close the page only when necessary ends up adding to the latency of the next request. In such a case, it is best to close the page as soon as possible~\cite{RBPP}. Accordingly, the close-page management policy (C$_{PM}$) immediately closes a row after accessing it.
	
	The C$_{PM}$ is not free of trade offs. It suffers excess delays and wasted power when there is some locality in the interleaved access streams. 
    Adaptive page policies such as the open-adaptive (OA$_{PM}$) and close-adaptive (CA$_{PM}$) have been proposed to improve upon O$_{PM}$ and C$_{PM}$. The OA$_{PM}$ policy closes the row only when: 1)~there are no more pending requests in the controller's queue that would hit in the open row, and 2)~there are pending requests to another row. The CA$_{PM}$ policy closes a row as soon as there are no pending requests that would hit in the same row. Thus, OA$_{PM}$ speculates that near future requests are likely to be hits. 
    
   Neither policy is always best. Accordingly, later work proposes switching between the two on-the-fly.
	Awasthi \textit{et al.} introduced the Access-Based Page Policy (ABPP) for multi-core systems \cite{ABPP}. ABPP assumes that a row will receive the same number of hits as the last time it was activated. The implementation uses per-bank tables recording the most recently accessed rows and the number of hits they received last time. The tables are used  to predict how long a row should stay open and are dynamically updated as the program executes. In the absence of a table entry, a row stays open until a conflict forces it to close.
	
	Shen \textit{et al.} proposed the Row-Based Page Policy (RBPP) which has lower hardware overhead than ABPP~\cite{RBPP}. RBPP exploits the observation that in a short period of time most of the memory requests are for a small number of rows even in a many-core system with different applications running on each core. RBPP uses a few most-accessed-row registers (MARR) per bank recording the number of hits received by recently accessed rows that have received at least one hit. 
	
    Timer-based page closure precedes RBBP and ABBP~\cite{adaptiveIdleTimer,dynamicIdleTimer} and predicts how long a row should stay open. 
    The various timer-based policies differ in the granularity of the timer, e.g., some maintain a timer per bank while others maintain a global timer. Other approaches used a branch prediction-like two-level predictor to decide whether to close an open row-buffer~\cite{prediction_pp:Xu,history_pp}. We opted not to include these proposals in our study since they were proposed for single core systems and would have to be adapted for many-core systems, moreover, RBPP and ABPP outperformed them~\cite{RBPP}.
	
	\section{Methodology} \label{Methodology}
	\subsection{Workloads}
	Our study focuses on the scale-out server workloads represented by the CloudSuite benchmark suite~\cite{clearing_clouds:babak}. CloudSuite includes Data Serving, MapReduce, SAT Solver, Web Frontend, Web Search and Media Streaming applications. To evaluate and compare these emerging workloads to traditional server workloads, we extended our experiments to include two traditional workload categories: 1) Transactional workloads including SPECweb99 and TPC-C. We ran TPC-C on two commercial database management systems. 2) Decision support workloads represented by three TPC-H queries; Q2, Q6 and Q17. The three queries cover a wide range of select-intensive, join-intensive and select-join queries. 
	
	For all the workloads, we use the same benchmarks configurations used by Ferdman \textit{et al.}~\cite{clearing_clouds:babak}. Table \ref{workloads} reports the workloads, categories and acronyms we use in the rest of this paper. 
	\begin{table}
		\centering
		\caption{Categorized Workloads and Abbreviations}
		\label{workloads}
		\begin{tabularx}{\linewidth}{l|X|l|X} \hline
			\textbf{Category}&\textbf{Category acronym}&\textbf{Workload}&\textbf{Acronym}\\ \hline \hline
			
			\multirow{6}{*}{Scale-out}& \multirow{6}{*}{SCO$_W$} & Data Serving & DS \\ \hhline{~~--} 
			&& MapReduce & MR\\ \hhline{~~--}
			&& SAT Solver & SS\\ \hhline{~~--}
			&& Web Frontend & WF\\ \hhline{~~--}
			&& Web Search & WS\\ \hhline{~~--}
			&& Media Streaming & MS\\ \Xhline{1.2pt}
			&\multirow{3}{*}{TRS$_W$}& SPECweb99 & WSPEC99\\ \hhline{~~--}
			Transactional& &TPC-C1 (vendor A) & TPC-C1\\ \hhline{~~--}
			& &TPC-C2 (vendor B)& TPC-C2\\ \Xhline{1.1pt}
			&\multirow{3}{*}{DSP$_W$}& TPC-H Q2 & TPCH-Q2\\ \hhline{~~--}
			Decision Support & &TPC-H Q6 & TPCH-Q6\\ \hhline{~~--}
			& &TPC-H Q17 & TPCH-Q17\\ 
			
			\hline\end{tabularx}
	\end{table}
	
	\subsection{Simulation}
	This study used the Virtutech Simics functional simulator to model a system of 16-core chip-multiprocessor (CMP) unless otherwise noted. The GEMS simulator~\cite{GEMS:Martin} was used to extend Simics with an on-chip network timing model and a Ruby-based full-memory hierarchy including the memory controller and off-chip DRAM timing models. The cores run the SPARC v9 ISA.
	
	We follow the SimFlex multiprocessor sampling technique for our simulation experiments~\cite{SimFlex}. The samples are taken over a 10-second simulated interval of each application. Each sample ran for six billion user-level instructions where the first one billion user-level instructions were used to warm up the system, e.g., the caches, memory queues, network buffers, and interconnects. Statistics were collected for the subsequent five billion user-level instructions. TPC-H queries Q2 and Q17 were run to completion for a total run length of roughly 2.5 billion user-level instructions, one billion of which was used to warm up. We simulate both user-level and operating system-level instructions but we use the user-level instruction count divided by the total simulated cycles count as an indicator of the overall system performance; a metric found by Wenisch \textit{et al.} to accurately represent system throughput~\cite{SimFlex}. Measured performance metrics include committed user-level instructions per cycle (user IPC), average memory access latency, DRAM row-buffer hit rate, LLC misses per kilo instructions (L2 MPKI), and memory bandwidth utilization. The Web Frontend benchmark uses only 8-cores in the configuration that was available to us.
	
	\subsection{Baseline System Configuration}
	The baseline system is a 16-core in-order CMP with only two levels of on-chip caches based on the scale-out processor design recommendations proposed by Lotfi-Kamran \textit{et al.}~\cite{scale_outProc}. In that study, pods of 16-32 in-order cores were found to achieve the highest performance density for scale-out workloads. The chip features a modestly sized  4MB LLC to capture the instruction working set and shared OS data. Table \ref{baseline_sys} details the architectural configurations, and Section \ref{channels_study} explains the address mapping abbreviation listed.
	
	\begin{table}
		\centering
		\caption{Baseline System Configuration}
		\label{baseline_sys}
		\begin{tabularx}{\linewidth}{|l|X|} \hline
			CMP Organization &16-core Scale-Out Processor pod\\ \hline
			Core & In-order @ 2GHz\\ \hline
			L1-I/D caches & 32KB each, 64B blocks, 2-way\\ \hline
			Shared L2 cache & 4MB, unified, 16-way, 64B blocks, 4 banks \\ \hline
			Interconnect & 16x4 crossbar\\ \hline
			Memory Controller & FR-FCFS scheduling, open-adaptive page policy, 1-channel, 11.9GB/s bandwidth, RoRaBaCoCh address mapping\\ \hline
			Off-chip DRAM & 32-64GB, DDR3-1600 (800MHz), 2 ranks, 8 banks per rank, 8KB row-buffer \\ \hline
			$t_{CAS}$-$t_{RCD}$-$t_{RP}$-$t_{RAS}$ & 11-11-11-28\\
			$t_{RC}$-$t_{WR}$-$t_{WTR}$-$t_{RTP}$ & 39-12-6-6\\
			$t_{RRD}$-$t_{FAW}$ (in cycles) & 5-24\\
			\hline\end{tabularx}
	\end{table}

	\section{Results} \label{results}
	Section \ref{Memory Scheduling Study} compares the memory scheduling techniques in terms of overall system performance, average memory access latency and row-buffer hit rate. The observations in Section \ref{row_buffer_reuse} motivate the study of Section \ref{page_mgnt_study} which identifies the efficiency of DRAM page management policies in predicting when to close a row-buffer and when to keep it open. Section \ref{channels_study} investigates the effectiveness of using multi-channel memory controllers. 
    
	\begin{table}
		\centering
		\caption{Scheduling Algorithms Configurations}
		\label{schedulers_configs}
		\begin{tabularx}{\linewidth}{X|l|X} \hline
			\textbf{Algorithm}&\textbf{Parameter}&\textbf{Value}\\ \hline \hline
			PAR-BS&Batching-Cap&5\\
			\Xhline{1.02pt}
			
			\multirow{3}{*}{ATLAS} & Quantum & 10M cycles\\ \hhline{~--} 
			& $\alpha$ (bias to current quantum) & 0.875\\ \hhline{~--}
			& Starvation threshold & 50K cycles\\ \Xhline{1.05pt}
			
			\multirow{6}{*}{RL} & \# of Q-value tables& 32\\ \hhline{~--}
			& Q-value table size&256 Q-values\\ \hhline{~--}
			& $\alpha$ (learning rate)& 0.1\\ \hhline{~--}
			& $\gamma$ (discount rate) & 0.95\\ \hhline{~--}
			& $\epsilon$ (random action probability)&0.05\\ \hhline{~--}
			& Starvation threshold & 10K cycles\\
			\hline\end{tabularx}
	\end{table}
    
	\subsection{Memory Scheduling Study} \label{Memory Scheduling Study}
	We evaluate the baseline FR-FCFS scheduling algorithm \cite{FR_FCFS:Rixner,FR_FCFS:Rixner2004,FR_FCFS:patent} as well as the state-of-the-art memory scheduling techniques PAR-BS \cite{PAR_BS:Mutlu}, ATLAS \cite{ATLAS:Mutlu} and RL \cite{RL:Mutlu}. We also evaluate the simplest algorithm FCFS\_banks. Results are normalized to the baseline FR-FCFS unless otherwise stated. Table \ref{schedulers_configs} shows the configurations used for PAR-BS, ATLAS and RL.
	
	\subsubsection{Performance}
	
	\begin{figure}
		\centering
		\includegraphics[width=3.35in,height=2in]{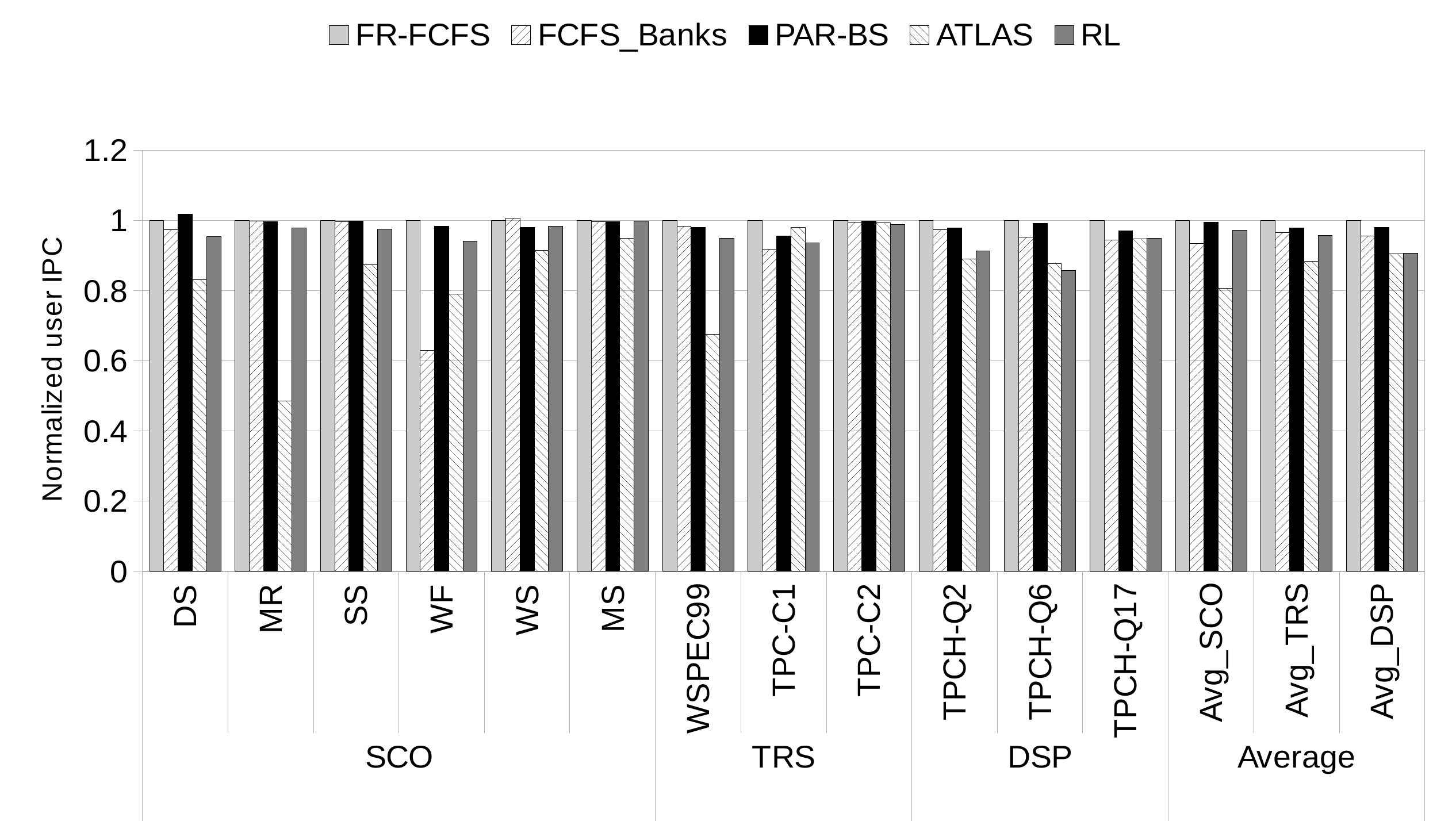}
		\caption{User IPC normalized to FR-FCFS.}
		\label{fig:op_ipcFRFCFS}
	\end{figure}
	
	Figure \ref{fig:op_ipcFRFCFS} shows the user IPC normalized to FR-FCFS. The results show that, under server workloads and especially SCO$_W$ applications, FR-FCFS outperforms PAR-BS, ATLAS and RL. The only exceptions are TPC-C2 and Media Streaming where the aforementioned policies perform as well as FR-FCFS. 
	
	Performance with ATLAS is lower more so for SCO$_W$ that suffers a 20\% average drop in performance compared to FR-FCFS. ATLAS with its long quantum period of 10 million cycles causes some cores to be unfairly given lower priority for long periods. For example, in MapReduce and Web Frontend some cores had 50\% lower IPC than others, resulting in overall performance degradation of 52\% and 21\% respectively. By comparison, the lowest per core IPC with FR-FCFS is within 85\% of the highest per core IPC for these two applications. Large IPC disparity across cores is also responsible for the 33\% performance loss of SPECweb99. On average, ATLAS achieves average performance that is 20\%, 12\% and 10\% lower for SCO$_W$, TRS$_W$ and DSP$_W$ respectively.
	
	RL also performs worse than FR-FCFS, more so for DSP$_W$ where the loss is 10\%. DSP$_W$'s access patterns tend to be more random than that of conventional desktop and scientific applications and thus challenge the RL exploration process which introduces overheads if activated frequently enough. 
	
	\begin{figure}
		\centering
		\includegraphics[width=3.35in,height=2in]{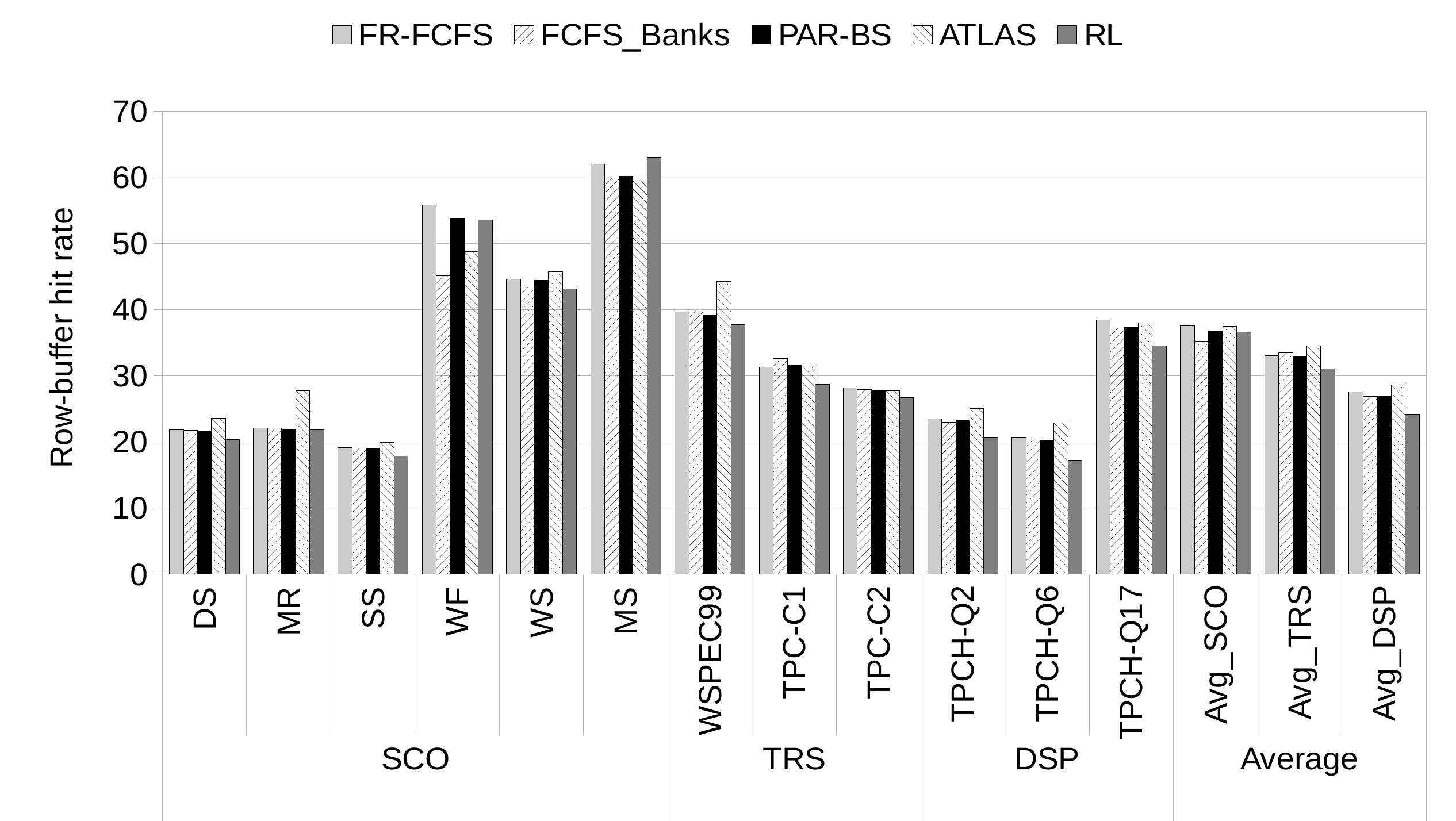}
		\caption{Row-buffer hit rate.}
		\label{fig:op_rbhits}
	\end{figure}
	
\begin{figure}
		\centering
		\includegraphics[width=3.35in,height=2in]{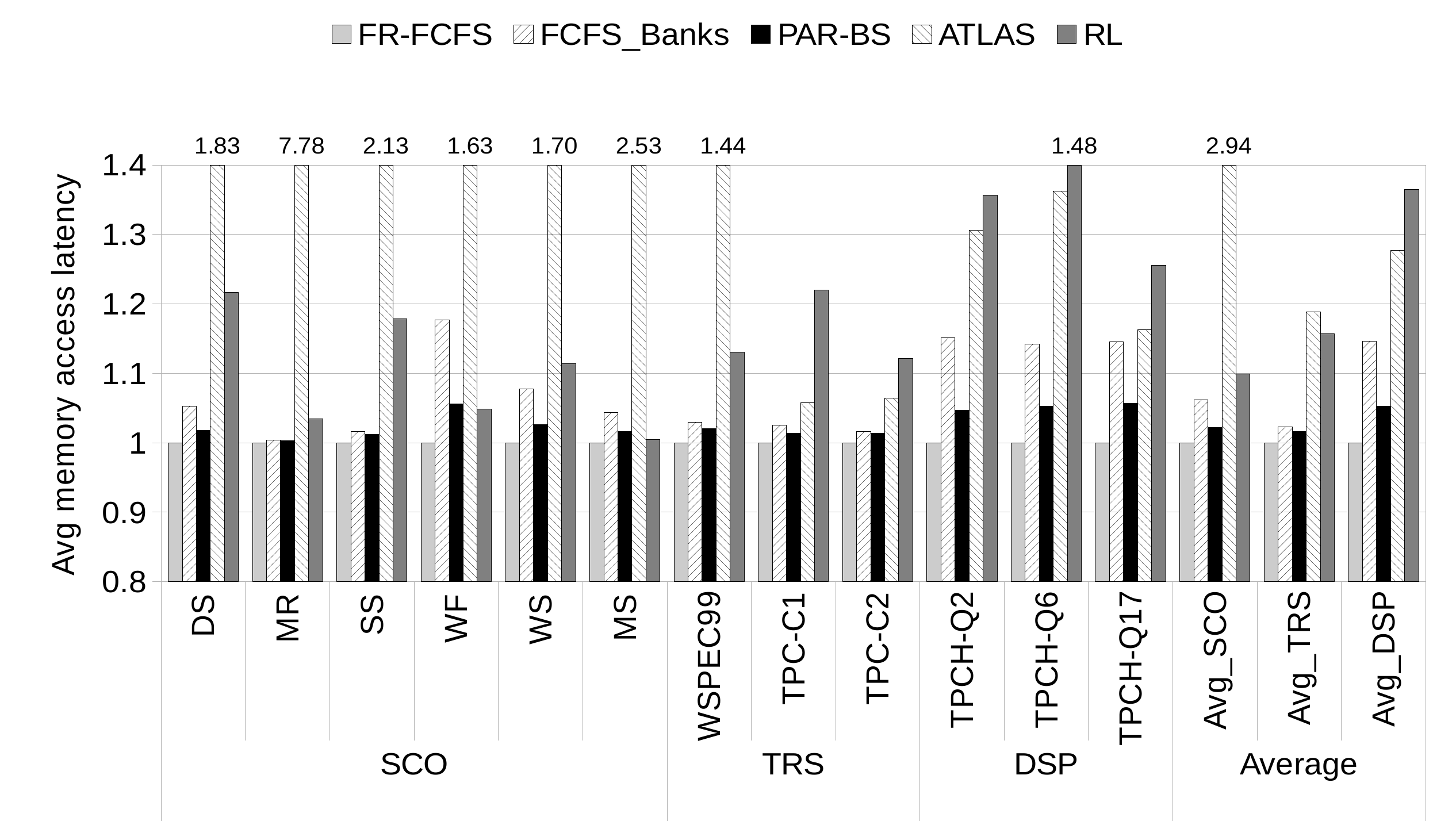}
		\caption{Average memory access latency normalized to FR-FCFS.}
		\label{fig:op_memAccFRFCFS}
	\end{figure}

	The simpler FCFS\_Banks performs closer to FR-FCFS than the other policies. FCFS\_Banks' average performance is within 6\%, 3\% and 4\% for SCO$_W$, TRS$_W$ and DSP$_W$ respectively. For five out of the six SCO$_W$ workloads, FCFS\_Banks matches the performance of FR-FCFS. This can be explained by~Figure \ref{fig:op_rbhits} which shows that the row-buffer hit rate changes by only -4\%, +1\% and -2\% for SCO$_W$, TRS$_W$ and DSP$_W$ respectively with FCFS\_Banks. FR-FCFS' benefit over FCFS\_Banks is that it favors row hits. For these workloads, the cores are either not concurrently competing for the same bank or they access the same row in the same bank at the same time. Accordingly, there is little benefit from reordering the requests heading to the same bank. Web Frontend is the exception as its performance is 37\%  lower with FCFS\_Banks. This workload exhibits a row-buffer hit rate of 55\% with FR-FCFS which drops to 45\% with FCFS\_Banks leading to a 17\% increase in average memory access latency as ~Figure \ref{fig:op_rbhits} and~Figure \ref{fig:op_memAccFRFCFS} show. 
	Compared to FR-FCFS, FCFC\_Banks does not need to scan all the request queues every cycle searching for a row hit to promote. As a result, it is simpler to implement and would require less energy. However, overall system energy may increase unless performance stays the same.
	
	\subsubsection{Memory Access Latency Sensitivity} \label{mem_acc_latency}
	The average memory access latency in~Figure \ref{fig:op_memAccFRFCFS} correlates with the changes in user IPC. However, the  sensitivity of performance to memory access latency differs across workloads.
    For example, DSP$_W$ is less sensitive to average memory latency than the Web Frontend workload; while DSP$_W$ suffers a 15\% increase in memory access latency with FCFS\_Banks this translates to a 4\% reduction in IPC. DSP$_W$ exhibits some memory level parallelism
    which hides most of the increase in memory access latency under FCFS\_Banks.
	
	\begin{figure}
		\centering
		\includegraphics[width=3.35in,height=2in]{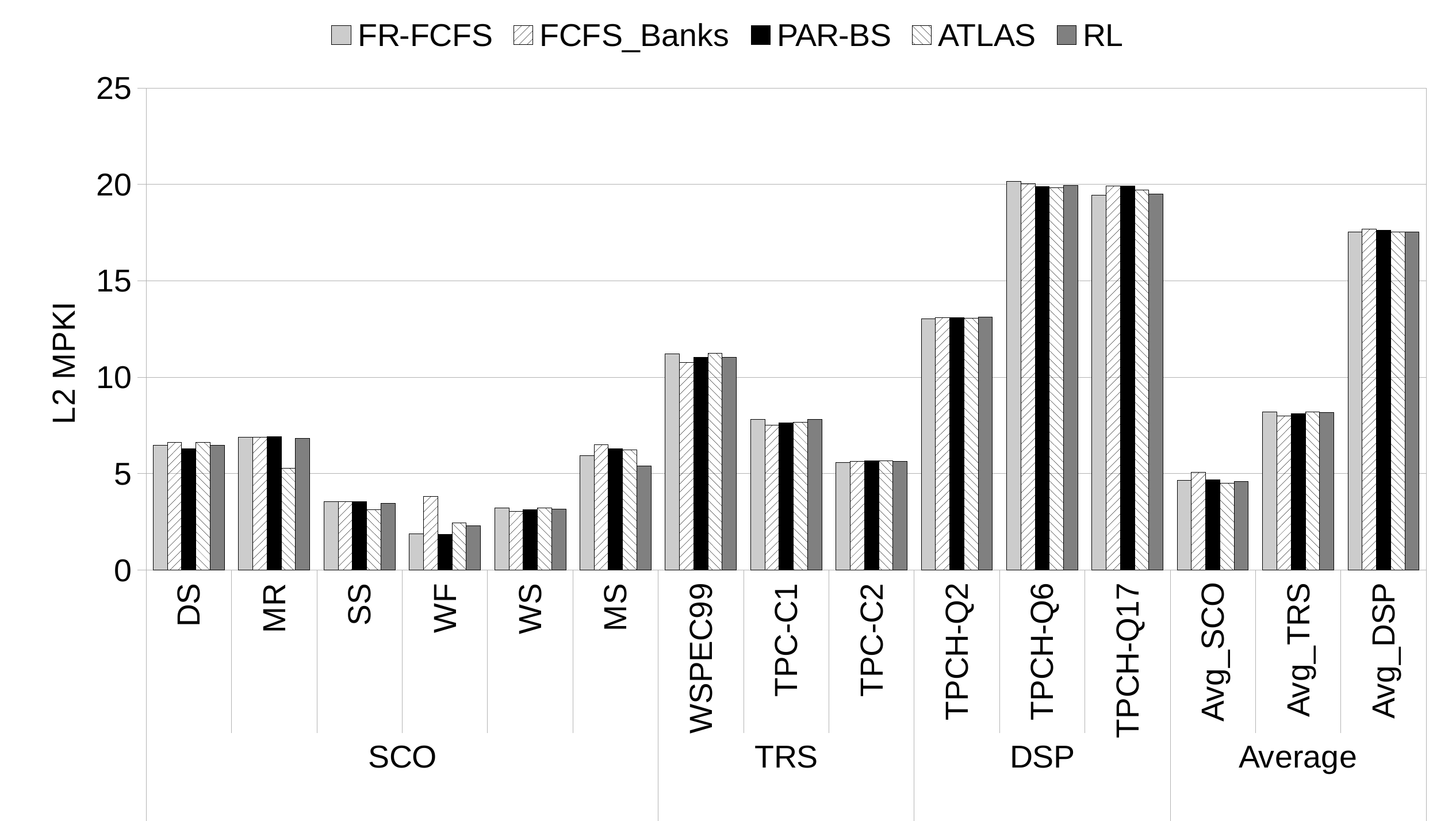}
		\caption{L2 Misses per kilo user instructions (MPKI).}
		\label{fig:op_MPKI}
	\end{figure}
	
	ATLAS suffers a significant increase in memory access latency which is more pronounced for SCO$_W$ at 2.94x on average and up to 7.78x for MapReduce. This increase in memory latency translates into a 20\% performance loss on average. 
    DSP$_W$ is also negatively impacted by ATLAS' long-term ranking scheme and RL's exploratory approach leading to 28\% and 37\% longer memory access latency respectively.
	
	The MPKI measurements in Figure \ref{fig:op_MPKI} indicate that SCO$_W$ and TRS$_W$ have relatively low memory intensity exhibiting an average L2 MPKI of 5 and 8 respectively. DSP$_W$ exhibits a higher average MPKI of around 18. This corroborates the different off-chip memory bandwidth demands of scale-out workloads reported in earlier work~\cite{clearing_clouds:babak}.
	
	\subsubsection{Requests Queue Length}
	\begin{figure}
		\centering
		\includegraphics[width=3.35in,height=2in]{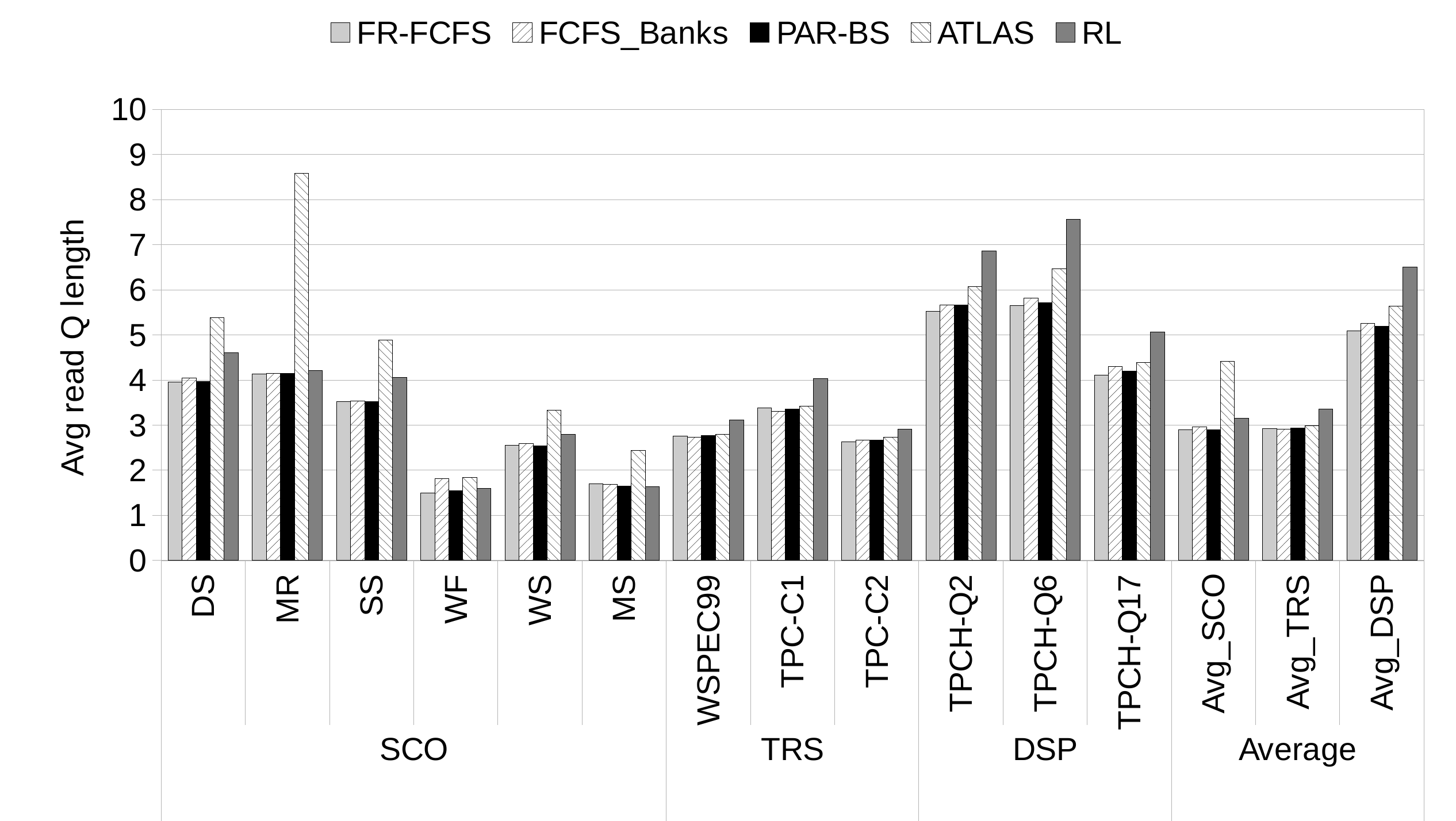}
		\caption{Average read queue length.}
		\label{fig:op_rq}
	\end{figure}
\begin{figure}
		\centering
		\includegraphics[width=3.35in,height=2in]{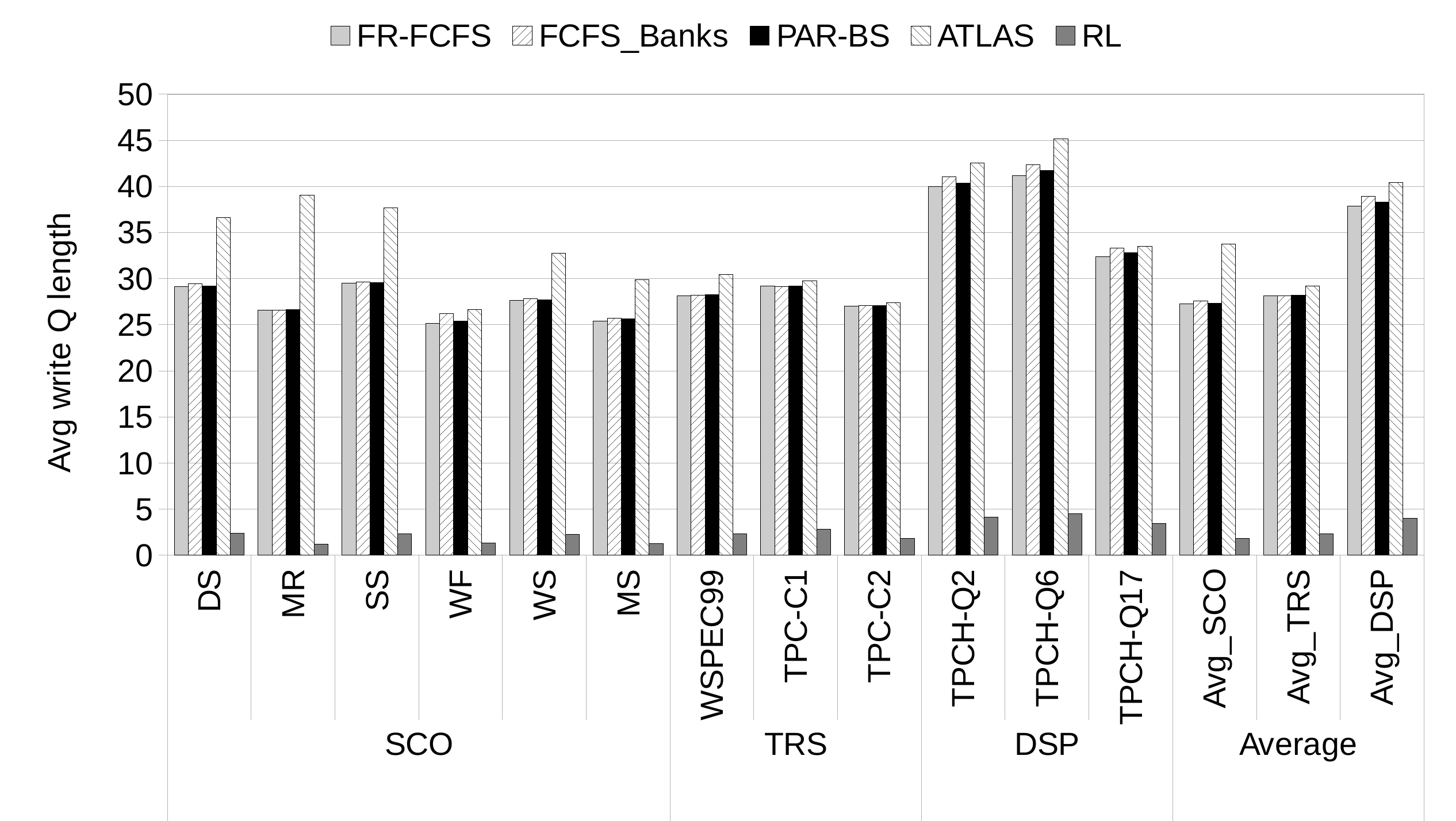}
		\caption{Average write queue length.}
		\label{fig:op_wq}
	\end{figure}

	The memory controller's average read and write queue occupancies (length) are shown in~Figure \ref{fig:op_rq} and~Figure \ref{fig:op_wq} respectively. 
    All memory scheduling techniques never needed more than a 10-entry read queue and a 50-entry write queue. On average, DSP$_W$ is more demanding than SCO$_W$ with MapReduce under ATLAS being the exception. MapReduce's behavior is due to the 7.78x increase in memory access latency as discussed in Section \ref{mem_acc_latency}. The observed queue lengths are far lower than those used in previous work~\cite{memory_design_book,gpu_mem_q_depth,mobile_mem_q_depth}. This may be due to the relatively simple in-order cores used here. 
	
	RL exhibits noticeably lower write queue lengths compared to the rest. The other techniques are switching from a read phase to a write phase only when the write queue length is above a certain threshold in order to reduce how often the data bus direction switching penalty is incurred.
RL considers both reads and writes when it selects the memory request to serve next and builds its decision based on the optimum strategy it learned so far. This gives the memory controller the freedom to serve write requests whenever it can steal a few cycles between critical memory reads.
	\begin{figure}
		\centering
		\includegraphics[width=3.35in,height=2in]{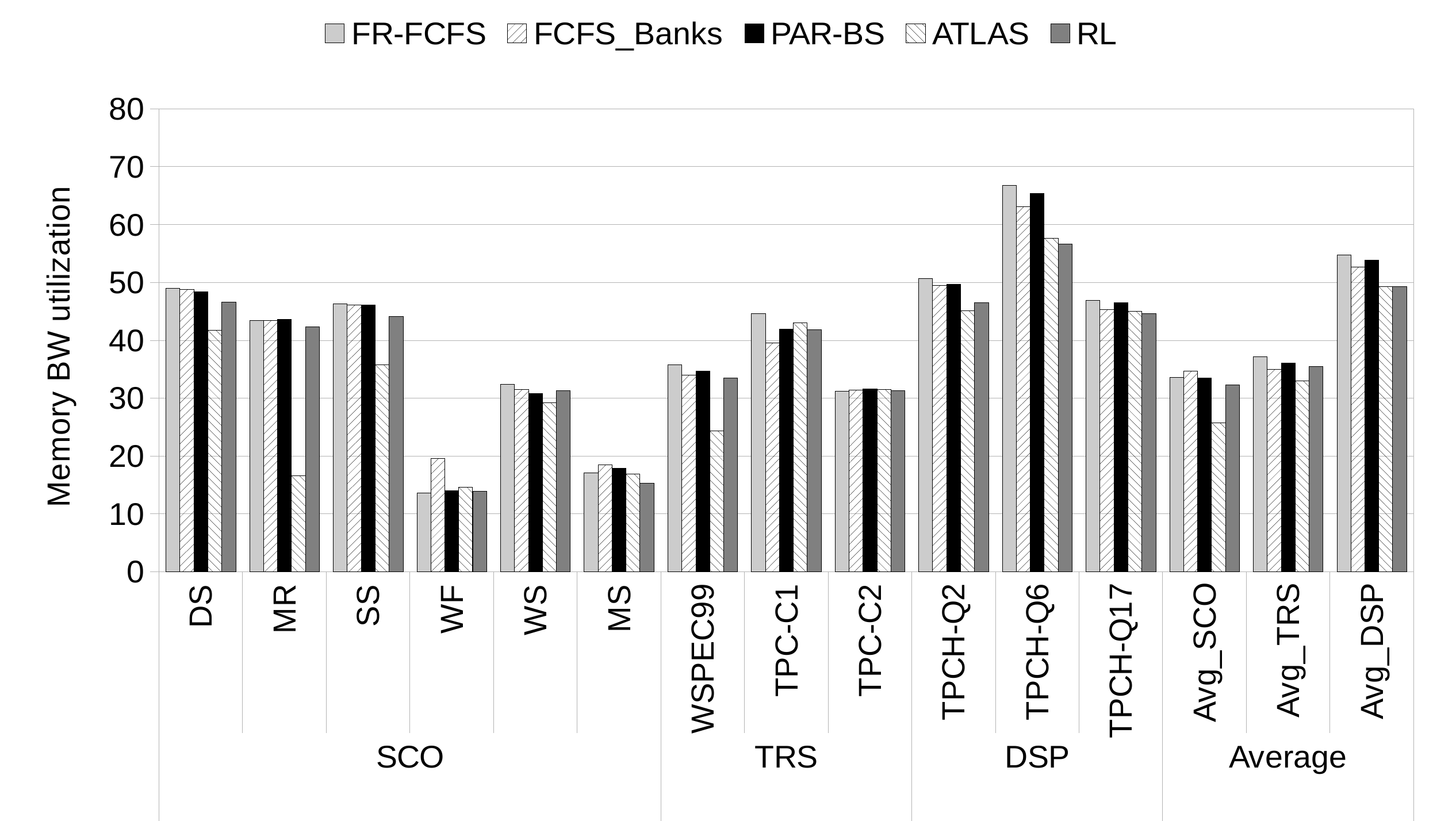}
		\caption{Memory bandwidth utilization.}
		\label{fig:op_bwAbs}
	\end{figure}
	
	\subsubsection{Off-Chip Bandwidth Utilization}

Figure \ref{fig:op_bwAbs}~shows the off-chip bandwidth utilization per workload. Utilization under SCO$_W$ ranges from 14\% and up to 50\% of the available peak bandwidth with an average of 34\%. TRS$_W$ exhibits a similar average while DSP$_W$  has higher off-chip access demands with an average utilization of 54\%. The measured average memory bandwidth utilization motivates the study in Section \ref{channels_study} that considers the impact of introducing additional memory channels.  
	
	\subsubsection{Summary of Memory Scheduling Study Results} 
	
	A technique as simple as FR-FCFS has proven best for the workloads studied under the pod-based in-order processor design proposed by Lotfi-Kamran \textit{et al.}~\cite{scale_outProc}. For the server workloads, and more so for scale-out workloads, FR-FCFS outperformed the other state-of-the-art memory scheduling techniques. These techniques were designed to outperform FR-FCFS for desktop and parallel applications such as SPEC CPU2006~\cite{SPEC2006:Henning} and SPLASH-2~\cite{SPLASH:Woo} in environments where fairness among threads or cores is a concern, which is not the case here. 
    Scale-out workloads exhibit relatively low MLP and different access patterns compared to scientific applications.
    Moreover, the cores used here are relatively simple as advised by Lotfi-Kamran \textit{et al.} who found that those workloads do not exhibit similarly high ILP as SPEC and PARSEC do. Combined these explain the differences in relative performance across the various memory scheduling policies.
    Even a simple FCFS scheduler that exploits bank-level parallelism is within 6\% of FR-FCFS in all cases and matches its performance for most.
	
	The analysis showed that relatively short read and write queues are sufficient for these workloads primarily due to the relatively low MLP. Finally, off-chip memory bandwidth utilization was shown to be relatively low suggesting that adding additional memory channels may not be needed if performance is the only consideration.
	
	\subsection{Page Management Policies Study} \label{page_mgnt_study}
	\subsubsection{Activated Row-Buffer Reuse} \label{row_buffer_reuse}
	
	\begin{figure}
		\centering
		\includegraphics[width=3.35in,height=2in]{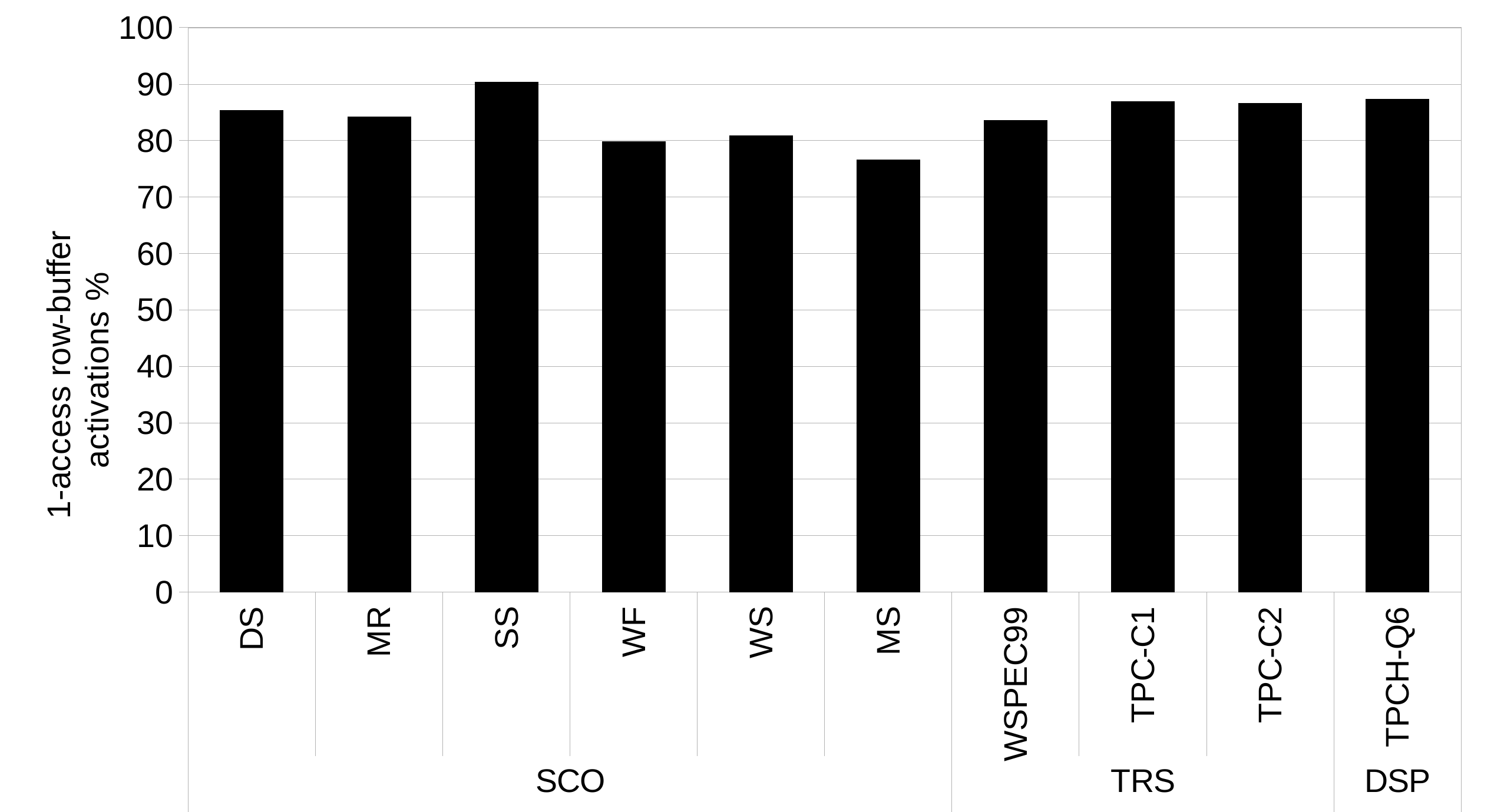}
		\caption{Percentage of single-access row-buffer activations under OA$_{PM}$.}
		\label{fig:op_1accrb}
	\end{figure}
	
	The baseline OA$_{PM}$ policy is often preferred for workloads that exhibit significant row-buffer locality. As  Figure \ref{fig:op_rbhits} shows, with this policy, the average row-buffer hit rates are relatively low at 37\%, 33\% and 27.5\% for SCO$_W$, TRS$_W$ and DSP$_W$ respectively. To reason about the access patterns of those workloads that might result in such row-buffer hit rates, we studied histograms of how frequently an activated row-buffer is reused before precharging it. 
    As Figure \ref{fig:op_1accrb} shows, all workloads access  77\%-90\% of their activated rows only once. Keeping a row open as long as possible thus will often prolong subsequent accesses that will have to first wait for the row to be closed.
    This observation suggests that using the CA$_{PM}$ policy might be better. 
	CA$_{PM}$ precharges the row-buffer directly after the column access if no more hits are waiting in the queue. 
    Having a high percentage of single-access row activations is not necessarily at odds with a high row-buffer hit rate. For example, in Media Streaming  76\% of the row activations observe only a single access while the remaining 24\% of the row activations experience a high number of hits.
    
	\begin{figure}
		\centering
		\includegraphics[width=3.35in,height=2in]{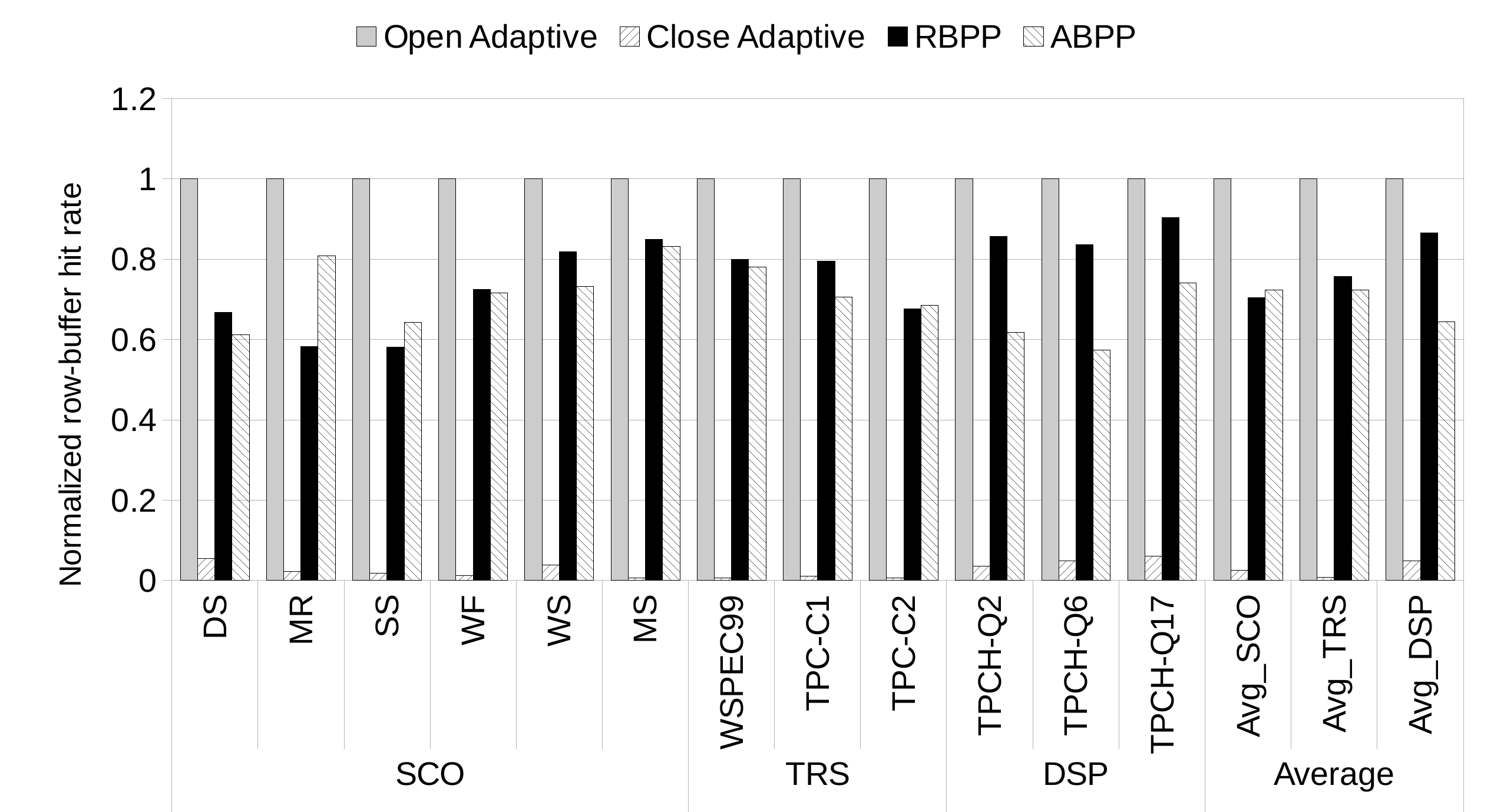}
		\caption{Row-buffer hit rate for different page management policies normalized to OA$_{PM}$.}
		\label{fig:op_rbhits_pagePolicies}
	\end{figure}
	
\begin{figure}
		\centering
		\includegraphics[width=3.35in,height=2in]{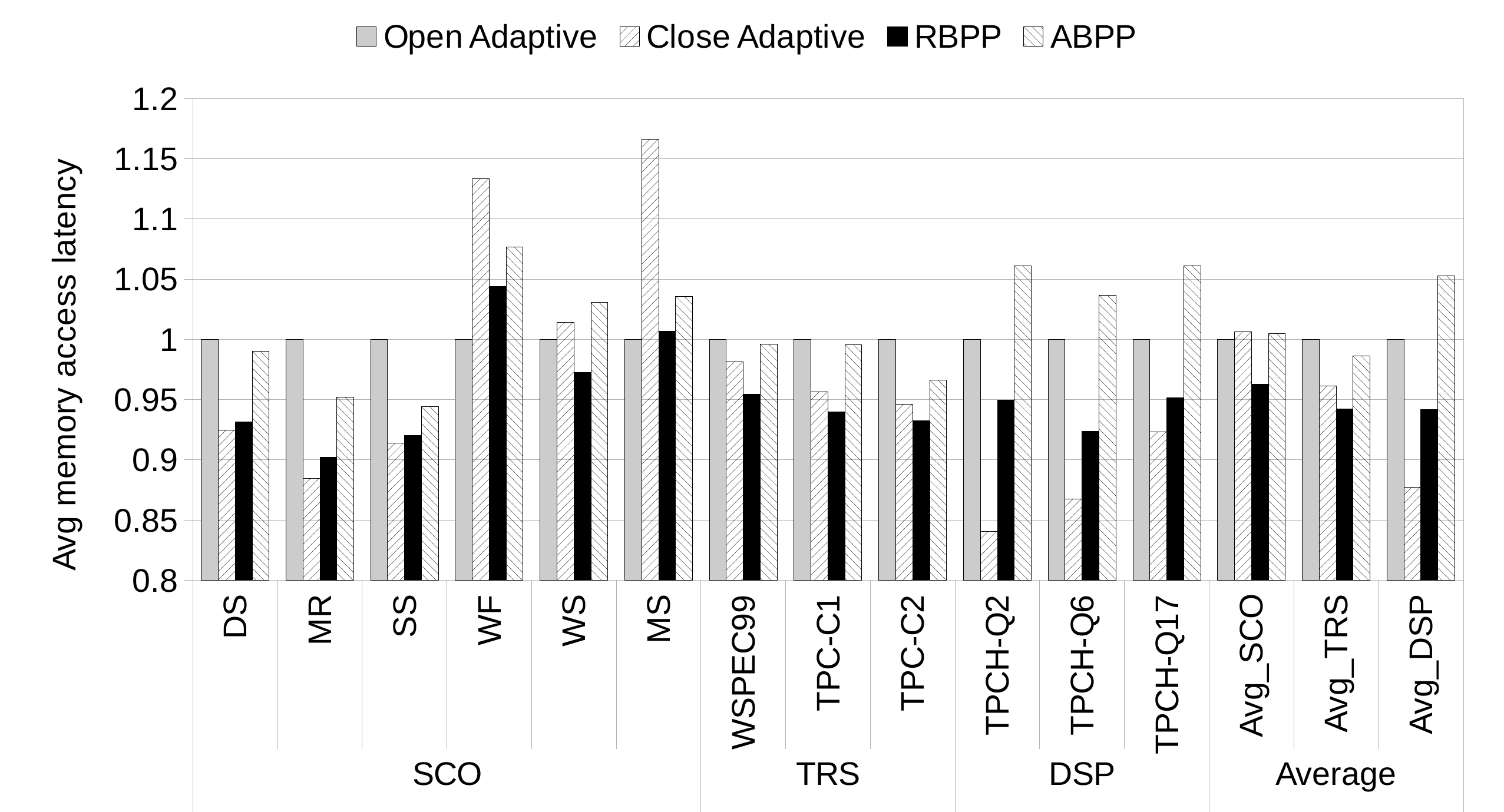}
		\caption{Average memory access latency for different page management policies normalized to OA$_{PM}$.}
		\label{fig:op_memAcc_pagePolicies}
	\end{figure}

\begin{figure}
		\centering
		\includegraphics[width=3.35in,height=2in]{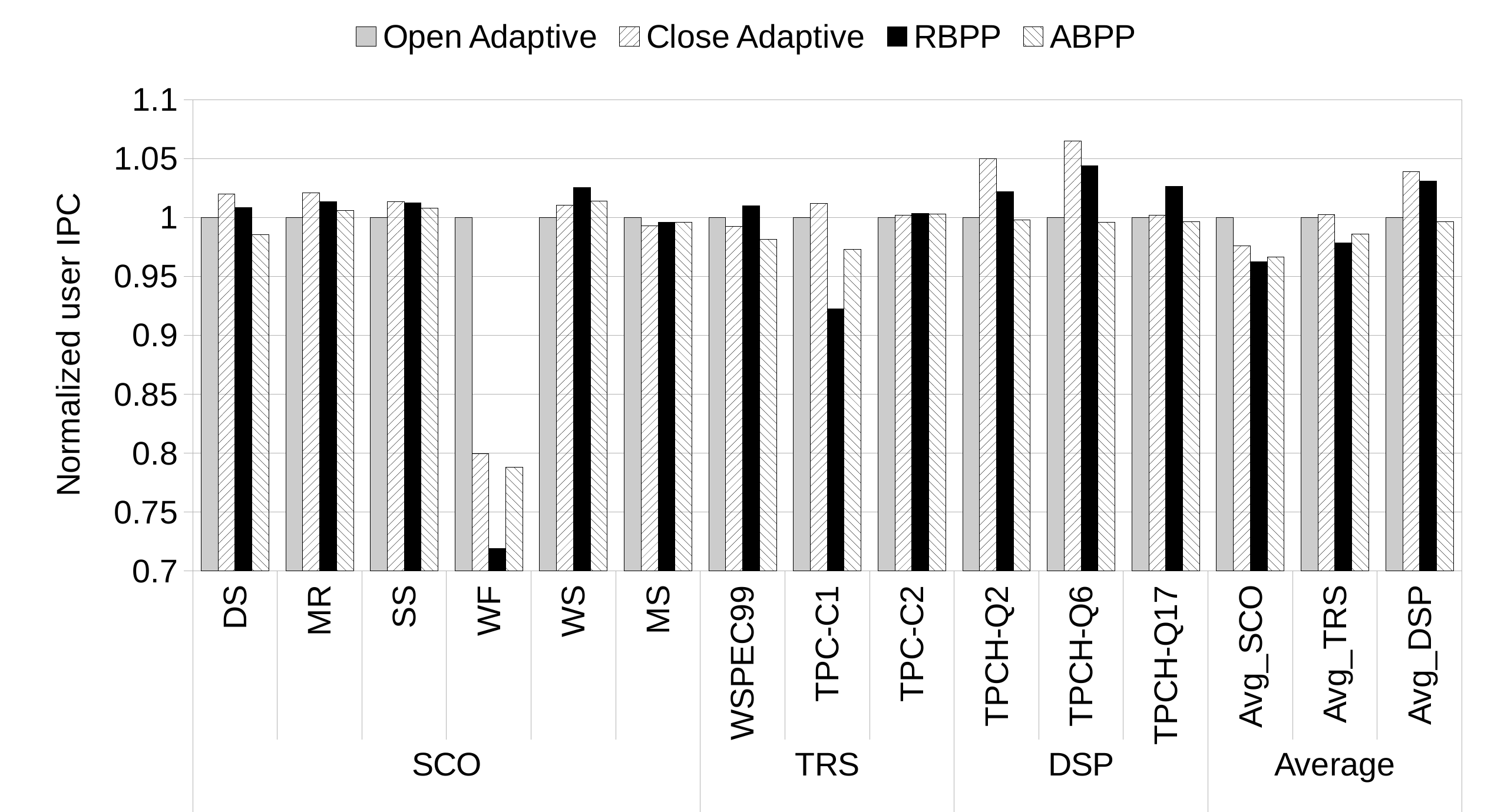}
		\caption{User IPC for different page management policies normalized to OA$_{PM}$.}
		\label{fig:op_userIPC_pagePolicies}
	\end{figure}

	Figure \ref{fig:op_rbhits_pagePolicies} shows that CA$_{PM}$ exhibits much lower row-buffer hit rates which are below 6\% for all workloads. This indicates that a significant fraction of the hits were due to the optimistic OA$_{PM}$. One would expect that this reduction in row-buffer hits will hurt performance and memory access latency. 
	In practice, the performance gained from closing the single-access rows under CA$_{PM}$ early compensates for the performance loss due to fewer row-buffer hits. Figure \ref{fig:op_memAcc_pagePolicies} shows that, under CA$_{PM}$, the average memory access latency did not change for SCO$_W$, while it decreased by 4\% and 13\% for TRS$_W$ and DSP$_W$. Performance was reduced by 2.5\% for SCO$_W$ but improved by 4\% for DSP$_W$ as Figure \ref{fig:op_userIPC_pagePolicies} shows. 
    Web Frontend and Media Streaming that exhibited the highest row-buffer hit rates under OA$_{PM}$, suffered a 15\% increase in access latency and Web Frontend lost 20\% of its performance. The relatively higher MPKI and MLP of Media Streaming limited its performance loss to 1\%. Data Serving, MapReduce and SAT Solver saw improved access latencies by 8\%, 12\% and 9\% and performance improvements in the range of 1\% to 2\%.

	\subsubsection{Predictive Page Management Policies} 
	The results of the previous section motivated us to study the behavior of state-of-the-art predictive page management policies.
This section compares RBPP~\cite{RBPP} and ABPP~\cite{ABPP} to CA$_{PM}$ and OA$_{PM}$ policies in terms of memory access latency, row-buffer hits preservation and user IPC. 
	
    As Figure \ref{fig:op_rbhits_pagePolicies} shows, RBPP is preserving 70\%, 75\% and 86\% of the row-buffer hits for SCO$_W$, TRS$_W$ and DSP$_W$ respectively. ABPP preserves generally less row-buffer hits more so for DSP$_W$. Figure \ref{fig:op_memAcc_pagePolicies} shows that DSP$_W$'s access latency is reduced by 6\% with RBPP, leading to a 3\% increase in user IPC in Figure \ref{fig:op_userIPC_pagePolicies}.  IPC for SCO$_W$ and TRS$_W$ degrades by 4\% and 2\% under RBPP. This correlates with the losses in row-buffer hits in Figure \ref{fig:op_rbhits_pagePolicies}.
	
	RBPP and ABPP policies favor capturing row-buffer hits over timely closure of single-access row activations. As a result, they do not avoid most of the penalty due to late closure of rows and they fail to capture some of the row hits when they prematurely close a row. 
    Overall performance is equal or slightly less than OA$_{PM}$. These policies were designed and tested for desktop applications such as SPEC CPU2006.
	
	\subsubsection{Summary of Page Management Policies Results}  
	This section found that scale-out workloads exhibit a  high percentage of single-access row activations. Thus, smarter page management policies are needed to timely close pages in order to achieve the following two goals: 1)~Capturing as many row hits as possible, and 2)~closing a page as early as necessary to avoid penalizing subsequent accesses to other rows. 
	
	\subsection{Multi-Channel Memory Systems Study} \label{channels_study}
	\begin{figure}
		\centering
		\includegraphics[width=3.35in,height=2in]{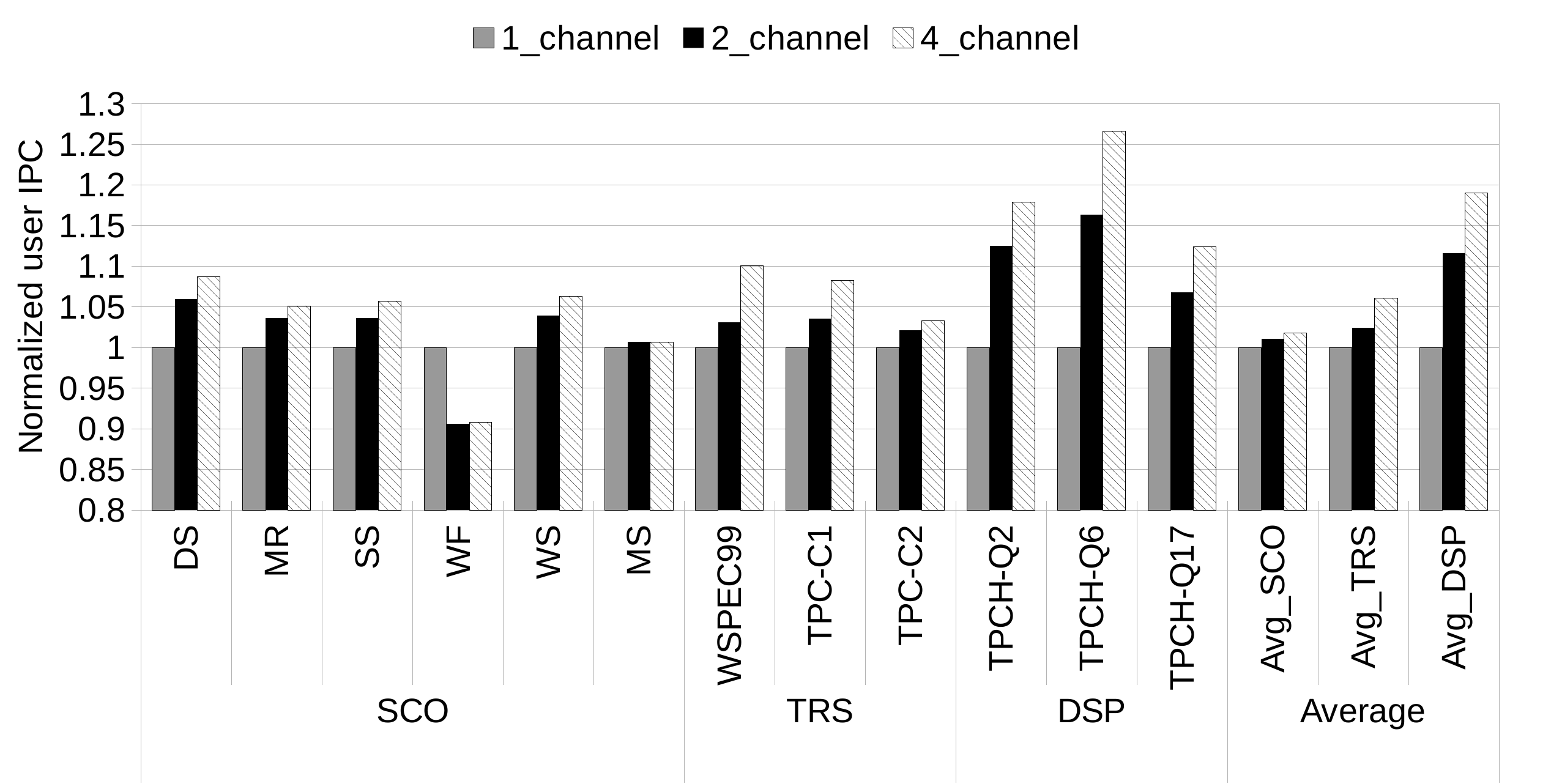}
		\caption{Normalized user IPC as the number of memory channels increases.}
		\label{fig:channels_userIPC}
	\end{figure}
	
	Modern server processors incorporate several memory channels~\cite{intel_xeon,amd_opteron}. This section considers the performance impact of using multi-channel memory controllers, a study motivated by the relatively low memory bandwidth utilization of the studied workloads. Specifically, this section studies the impact of integrating 2 and 4 memory channels in terms of performance, memory access latency and row-buffer hit rate. 
	The address mapping scheme, that is how physical addresses are mapped across the off-chip memory system, can impact overall performance. For this reason, we studied a number of address schemes that differ in which address bits they use to select the DRAM channel (Ch), column (Co), bank (Ba), rank (Ra) and row (Ro).
    The schemes studied are: RoRaBaCoCh (baseline mapping), RoRaBaChCo, RoRaChBaCo and RoChRaBaCo. In the interest of space, we report the results of the best performing scheme per workload and the averages. 
    The best performing scheme for each workload is reported in Section \ref{mapping_workloads}
    \begin{table}
    	\centering
    	\caption{The best performing multi-channel mapping scheme for each workload}
    	\label{mapping_workloads}
    	\begin{tabularx}{\linewidth}{X|X|X} \hline
    		\textbf{Workload}&\textbf{2-channel}&\textbf{4-channel}\\ \hline \hline
    		Data Serving & RoRaBaChCo &  RoRaChBaCo \\ \hline
    		MapReduce & RoRaChBaCo & RoChRaBaCo \\ \hline
    		SAT Solver & RoChRaBaCo	& RoRaChBaCo \\ \hline
    		Web Frontend & RoChRaBaCo	& RoRaBaCoCh \\ \hline
    		Web Search & RoRaChBaCo	& RoRaBaChCo \\ \hline
    		Media Streaming & RoChRaBaCo	& RoChRaBaCo \\ \hline
    		WSPEC99 & RoRaBaCoCh	& RoRaChBaCo \\ \hline
    		TPC-C1 & RoRaBaChCo	& RoChRaBaCo \\ \hline
    		TPC-C2 & RoRaChBaCo	& RoChRaBaCo \\ \hline
    		TPC-H Q2 & RoRaBaChCo	& RoChRaBaCo \\ \hline
    		TPC-H Q6 & RoRaChBaCo	& RoChRaBaCo \\ \hline
    		TPC-H Q17 & RoRaBaChCo	& RoChRaBaCo \\ 
    		\hline\end{tabularx}
    \end{table}

\begin{figure}
		\centering
		\includegraphics[width=3.35in,height=2in]{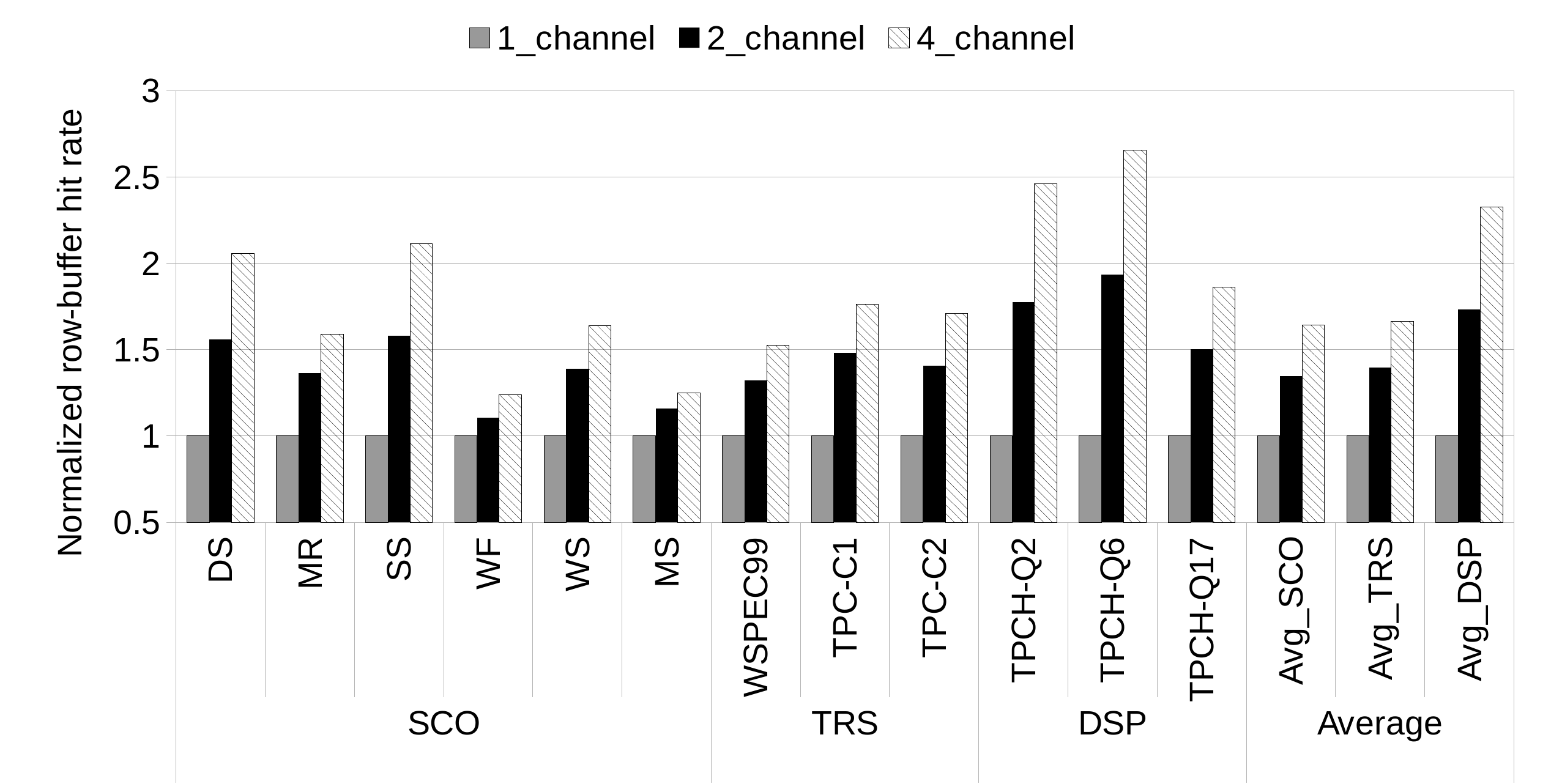}
		\caption{Normalized row-buffer hit rates as the number of memory channels increases.}
		\label{fig:channels_rbhits}
	\end{figure}

    The baseline scheme, RoRaBaCoCh, generally had the worst performance because accesses that would row hit in the 1-channel system may now map to a different channel; The scheme alternates successive cache blocks between the memory channels which means that sequential accesses do not map to the same DRAM row. 
     As shown in Figure \ref{fig:channels_rbhits}, RoRaBaChCo, RoRaChBaCo and RoChRaBaCo exhibit better row-buffer hit rates than the baseline system.
	Figure \ref{fig:channels_userIPC} shows that integrating more on-chip memory channels does not enhance SCO$_W$'s performance significantly and in some cases it hurts performance. The performance of Web Frontend dropped by 10\% and 9\% for the multi-channel systems. The highest gain of 6\% and 8.5\% for the 2- and the 4-channel configurations was observed for Data Serving. The average improvement for SCO$_W$ was below 1\% and 1.7\% for the 2- and 4-channels systems. 
    DSP$_W$ behaves differently and exhibits 11.5\% and 19\% average performance improvements for the 2- and 4-channel systems. Finally, TRS$_W$ also exhibits average performance improvements of 2.3\% and 6\% on the 2- and 4-channel systems.
	
	\begin{figure}
		\centering
		\includegraphics[width=3.35in,height=2in]{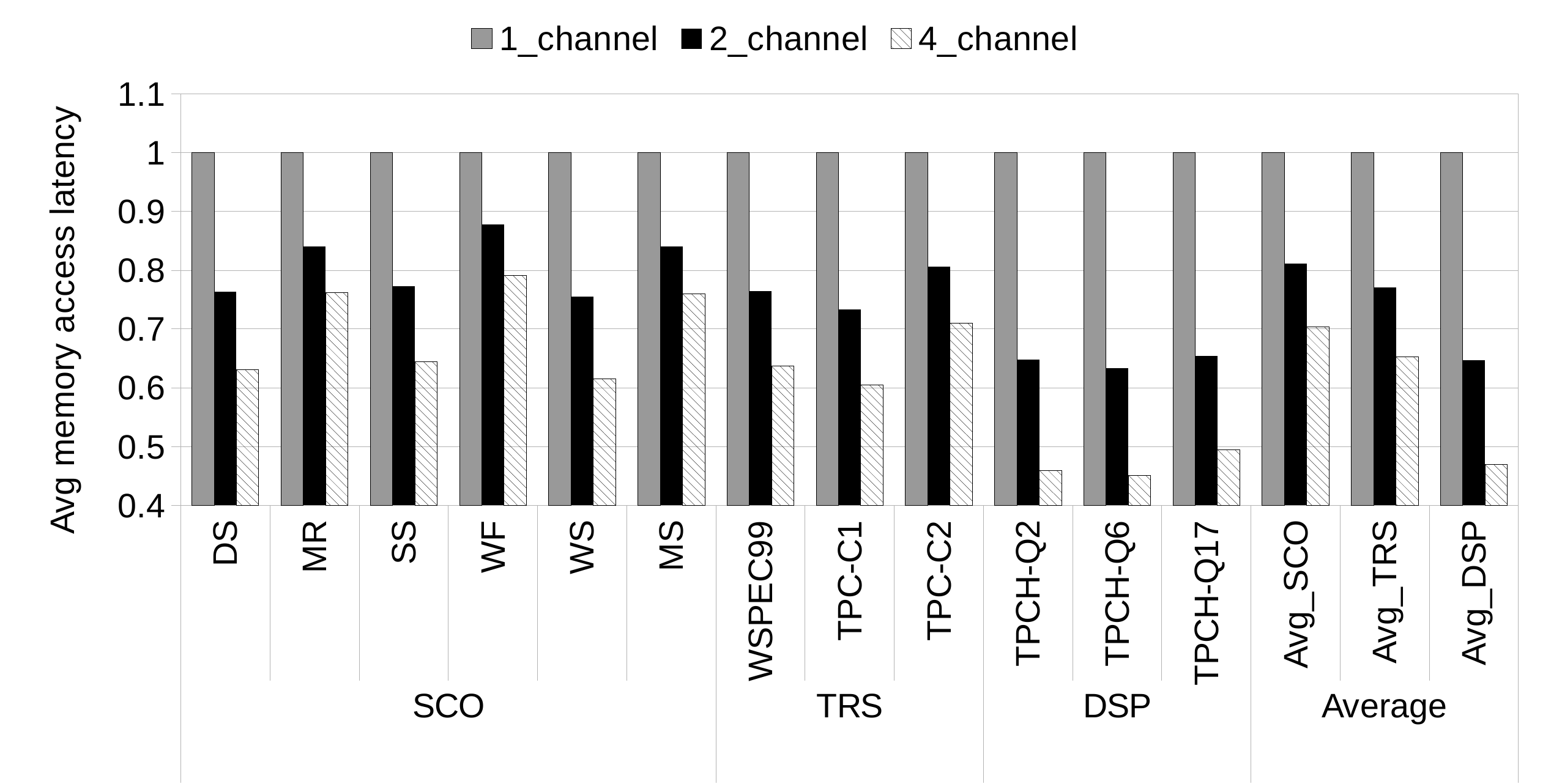}
		\caption{Normalized memory access latency as the number of memory channels increases.}
		\label{fig:channels_acc_latency}
	\end{figure}
	Figure \ref{fig:channels_rbhits} and Figure \ref{fig:channels_acc_latency} explain why performance did not improve for SCO$_W$ and TRS$_W$ in contrast to DSP$_W$. SCO$_W$ and TRS$_W$ improved the least in terms of row-buffer hit rates and memory access latencies. For both categories, the average row-buffer hit rate increased by 1.3x and 1.6x for the 2- and 4-channels systems. The average memory access latency of SCO$_W$ decreased to 81\% and 70\% of the baseline system. TRS$_W$ experiences similar reductions for the average memory access latency. Meanwhile, DSP$_W$'s average hit rates increased by 1.7x and 2.3x on the 2- and 4-channels systems respectively. Moreover, DSP$_W$'s memory access latency decreased to 64\% and 47\% of the baseline access latency. Along with the relatively higher MPKI of DSP$_W$, shown in Figure \ref{fig:op_MPKI}, this explains the difference in the IPC gains between DSP$_W$ and SCO$_W$ when multiple channels are used.
	
	Regardless of the improvements in row-buffer hits and memory access latency, Web Frontend's performance dropped by around 10\%. This workload exhibited an 11\% and 25\%  increase in the total number of memory accesses for the 2- and 4-channel systems. The extra memory references are mostly DMA/IO and atomic memory requests which hit in the row buffers reducing average latency. However, they increase contention and latency for critical accesses thus hurting performance. The reported user IPC, which is a metric of the user-level progress per cycle, is expected to go down due to more congestion from the extra accesses.
	
	\textbf{Conclusion:} Under the pod-based in-order processor design proposed by Lotfi-Kamran \textit{et al.} \cite{scale_outProc}, integrating multi-channel memory controllers on-chip with higher die area and power consumption does not improve the performance for scale-out workloads. 
    One channel can satisfy the low off-chip memory bandwidth demand of scale-out workloads  under the studied pod configuration. However, increasing core count would lead to a higher demand for off-chip bandwidth which could benefit from multiple channels. 
	
	\section{Limitations of this study}\label{limitations}
	The focus of this study is limited to the pod-based in-order scale-out processor design proposed by Lotfi-Kamran \textit{et al.} \cite{scale_outProc}. Although their study proposed an additional out-of-order scale-out processor design, we studied the in-order design as it was demonstrated to have higher performance density, i.e., throughput per unit area. Aggressive out-of-order designs might lead to different conclusions about how simple the memory scheduling technique should be and the needed off-chip memory bandwidth due to a potential increase in the MLP generated under such architectures. 
   
    Lotfi-Kamran \textit{et al.} assumed a 270 mm$^{2}$ die, which was estimated to fit three 32-core pods and six memory controllers. This work studies the memory system requirements for one such pod representative of a lower-end server processor. Future work, should consider higher pod counts. 

    The study did not include the TCM~\cite{TCM} memory controller policy which also targets fairness; experiments with ATLAS and PAR-BS showed that fairness is not an issue for scale-out workloads.
    
    The study includes a subset of the possible address mapping schemes and did not consider additional  permutation-based interleaving schemes. However, the results presented here have identified performance deficiencies which could guide such a future study.  
   
   This study is limited to reevaluating previous proposals for memory scheduling algorithms, page management policies and multi-channel memory controllers. While no novel solution is presented, this study provides directions for simplifying or improving these designs to better match the needs of emerging server workloads. Finally, this study focused solely on performance. Energy and power are equally important considerations. The performance results presented here can be useful in such a future study.
   
	\section{Related Work} \label{related work}
	Rixner \textit{et al.} introduced FR-FCFS scheduling techniques as well as other techniques that give preference to column access commands maximizing row-buffer hits and memory throughput \cite{FR_FCFS:Rixner,FR_FCFS:Rixner2004}. Rixner targeted web server workloads from SPECweb99 that exposed different memory access behavior than the currently wide spreading scale-out applications. We targeted different workloads and emphasized directions that could simplify memory controller design. 
	
	Natarajan \textit{et al.} studied the impact of several memory controller design aspects on server processor performance \cite{MC_Server:Natarajan}. The study investigated open-page vs. close-page policies, in-order vs. out-of-order memory requests scheduling  and memory ranks interleaving. However, the study was limited to shared-bus multi-processor architectures and was based on synthetic random address traffic. Our study covers wider aspects of memory controller design, includes state-of-the-art proposals for each and studies full applications. 
	
Barroso \textit{et al.} introduced Piranha \cite{Piranha} an early architecture that balanced complexity, performance and energy to better meet the demands of server applications.
PicoServer \cite{PicoServer:Kgil} is a simplified-core CMP design for server workloads that also favors relatively small on-chip caches. PicoServer relies on 3D die-stacking technology to provide low latency DRAM access. 
	
	Abts \textit{et al.} studied intelligent memory controller placement in many-core server processors~\cite{MC_Placement:Abts}. The study proposed diamond placement within the many-core tiles along with an enhanced routing algorithm for requests and replies, namely the class-based deterministic routing, to avoid the hot spots caused by the memory controllers. Our study is orthogonal to Abts's placement research as we investigate the microarchitecture of the memory controllers that better suits the needs of the emerging cloud workloads.
	
	Hardavellas \textit{et al.} proposed using CMPs for scale-out server workloads through exploiting heterogeneous architectures and dark silicon to power up only the application-suitable cores~\cite{dark_silicon:babak}. The work was extended by Ferdman \textit{et al.}~\cite{clearing_clouds:babak} where scale-out workloads were shown to behave differently than traditional server workloads. The comparisons showed that commodity server processors are over-provisioned in several ways. Building upon that study, Lotfi-Kamran \textit{et al.} introduced a pod-based CMP design that increases performance density~\cite{scale_outProc}. 

	\section{Conclusions} \label{conclusion}
	Previous work had characterized the behavior of scale-out workloads and its impact on core architecture and the on-chip memory hierarchy. 
	
	This work builds upon these past studies by also studying the off-chip memory access characteristics of scale-out workloads and their interaction with state-of-the-art memory controller policies. 
    The results showed that relatively simple memory scheduling algorithms worked best for these workloads outperforming other more advanced algorithms tuned to desktop workloads. The design of these advanced algorithms need to be revisited for scale-out workloads. We found that scale-out workloads exhibit poor row-buffer locality and, thus, better memory page management policies are needed to take advantage of the locality that exists while avoiding delaying the majority row-conflict requests. 
    
    Finally, additional memory channels did not significantly improve the performance for scale-out workloads. However, other scheduling policies, a different memory mapping, a different core architecture, or different data sets could have resulted in a different conclusion.

	\bibliographystyle{IEEEtran}
	\bibliography{IEEEabrv,references}

\end{document}